\documentclass[10pt,
							 twocolumn,
							 superscriptaddress,
							 english,
							 prl,
							 showpacs,
							 floatfix,
							 aps
							]{revtex4-1}
\usepackage[utf8]{inputenc}
\usepackage{amsmath}
\usepackage{amssymb}
\usepackage{graphicx}
\usepackage{xspace}

\makeatletter
\@ifundefined{textcolor}{}
{%
 \definecolor{BLACK}{gray}{0}
 \definecolor{WHITE}{gray}{1}
 \definecolor{RED}{rgb}{1,0,0}
 \definecolor{GREEN}{rgb}{0,1,0}
 \definecolor{BLUE}{rgb}{0,0,1}
 \definecolor{CYAN}{cmyk}{1,0,0,0}
 \definecolor{MAGENTA}{cmyk}{0,1,0,0}
 \definecolor{YELLOW}{cmyk}{0,0,1,0}
}

\usepackage[normalem]{ulem}

\usepackage{soul}
\usepackage{braket}
\usepackage[backref=none,
bookmarksnumbered=true,
bookmarks=true,
bookmarksopen=true,
colorlinks=true,
citecolor=blue,
linkcolor=blue,
anchorcolor=green,
urlcolor=blue,unicode=false]{hyperref}

\let\baraccent=\= 
\renewcommand{\=}[1]{\stackrel{#1}{=}} 

\newcommand{\unitspace}{~}

\newcommand{\didv}{\ensuremath{{\mathrm d}I/{\mathrm d}V}\xspace}

\newcommand{\Fig}[1]{Fig.\unitspace\ref{fig:#1}}
\newcommand{\Figure}[1]{Figure\unitspace\ref{fig:#1}}

\newcommand{\YSR}{YSR\xspace}
\newcommand{\udir}[1]{$\left<#1\right>$}
\newcommand{\dir}[1]{$\left[#1\right]$}

\DeclareMathOperator{\unm}{\unitspace\mathrm{nm}}

\DeclareMathOperator{\umV}{\unitspace\mathrm{mV}}

\DeclareMathOperator{\umeV}{\unitspace\mathrm{meV}}
\DeclareMathOperator{\umueV}{\unitspace\mathrm{\mu eV}}

\DeclareMathOperator{\upA}{\unitspace\mathrm{pA}}

\DeclareMathOperator{\uK}{\unitspace\mathrm{K}}

\DeclareMathOperator{\uAA}{\unitspace\mathrm{\AA{}}}

\makeatother

\usepackage{babel}

\newcommand{\abs}[1]{\left| #1 \right|} 

\begin{document}
\title{Wave-function hybridization in Yu-Shiba-Rusinov dimers
}

\author{Michael Ruby}
\affiliation{\mbox{Fachbereich Physik, Freie Universit\"at Berlin, 14195 Berlin, Germany}}

\author{Benjamin W. Heinrich}
\affiliation{\mbox{Fachbereich Physik, Freie Universit\"at Berlin, 14195 Berlin, Germany}}

\author{Yang Peng}
\affiliation{\mbox{Dahlem Center for Complex Quantum Systems and Fachbereich Physik, Freie Universit\"at Berlin, 14195 Berlin, Germany}}
\affiliation{Institute of Quantum Information and Matter and Department of Physics,California Institute of Technology, Pasadena, CA 91125, USA}
\affiliation{Walter Burke Institute for Theoretical Physics, California Institute of Technology, Pasadena, CA 91125, USA}

\author{Felix von Oppen}
\affiliation{\mbox{Dahlem Center for Complex Quantum Systems and Fachbereich Physik, Freie Universit\"at Berlin, 14195 Berlin, Germany}}

\author{Katharina J. Franke}
\affiliation{\mbox{Fachbereich Physik, Freie Universit\"at Berlin, 14195 Berlin, Germany}}
\date{\today}
\begin{abstract}
Magnetic adsorbates on superconductors induce local bound states within the superconducting gap. These Yu-Shiba-Rusinov (\YSR) states decay slowly away from the impurity compared to atomic orbitals, even in 3d bulk crystals. Here, we use scanning tunneling spectroscopy to investigate their hybridization between two nearby magnetic Mn adatoms on a superconducting Pb(001) surface. We observe that the hybridization leads to the formation of symmetric and antisymmetric combinations of \YSR states. We investigate how the structure of the dimer wave functions and the energy splitting depend on the shape of the underlying monomer orbitals and the orientation of the dimer with respect to the Pb lattice.
%
\end{abstract}
\pacs{%
			} 
\maketitle 


Magnetic adatoms on superconductors induce a local pair-breaking potential which binds Yu-Shiba-Rusinov (\YSR) states inside the superconducting energy gap~\cite{Yu, Shiba, Rusinov69}. The symmetry of the potential derives from the orbital symmetry of the spin-polarized states of the adsorbate ~\cite{Schrieffer67, Moca08, RubyMn16}. If the substrate imposes a sufficiently strong crystal field, the degeneracy of the adatom $d$ levels, and consequently also of the \YSR states, will be lifted ~\cite{RubyMn16, Choi16}. It was already predicted by Rusinov that the \YSR wave functions of two nearby adatoms hybridize and form bonding and antibonding combinations when the magnetic moments of the adatoms align ferromagnetically~\cite{Rusinov69}. Subsequent theoretical studies explored the spatial structure of the \YSR patterns~\cite{Flatte2000, Morr2003} and the phase diagram~\cite{Yao2014} for \YSR dimers. Many theoretical treatments assumed classical adatom spins with fixed alignment. More generally, additional energy scales such as Hund couplings, crystal fields, and magnetocrystalline anisotropies affect the interaction of the magnetic adatoms~\cite{Zitko11, Hatter15}. In quantum spin systems, Kondo screening also needs to be considered ~\cite{Hatter15, Franke11, Grove17}. 

\YSR states have considerable lateral extent away from the magnetic adatom~\cite{Yazdani97, Menard15}. This leads to wave-function hybridization and energy splitting of \YSR states in dimers of magnetic adatoms. 
Recent experimental studies observed these splittings for manganese (Mn) atoms on Pb(111), cobalt-phthalocyanine on NbSe$_2$, and chromium on $\beta$-Bi$_2$Pd~\cite{Ji08, Kezilebieke17, Choi2017BiPd}. While the latter two systems exhibit only a single \YSR resonance, Mn adatoms on Pb(111) show several crystal-field-split \YSR states~\cite{RubyMn16}. Starting with \cite{Ji08}, the earlier experiments already provided some indications of bonding and antibonding  \YSR states, but did not resolve how different \YSR states are affected by the coupling to a neighboring adatom and how the orbital nature of the \YSR states influences their hybridization.

Here, we present a  scanning tunneling microscopy and spectroscopy (STM/STS) study of dimers of Mn adatoms on Pb(001). The Mn adatoms adsorb in hollow sites with a square-pyramidal crystal field which governs the orbital wave functions of the \YSR states. We resolve symmetric and antisymmetric  combinations of the individual \YSR wave functions as well as a distinct distance- and angle-dependence of the  hybridization of the \YSR states. Our experimental study is complemented by a theoretical analysis of \YSR dimers which takes the orbital structure of the impurity states into account. 

For all measurements, we used a commercial SPECS JT-STM, which works under UHV conditions and at a base temperature of $1.2\uK$. The Pb(001) single crystal surface ($T_\mathrm{c}=7.2\uK$) was cleaned by cycles of Ne$^+$ ion sputtering and annealing until clean and atomically flat terraces were observed. Mn adatoms were evaporated onto the pre-cooled sample in the STM ($T<10\uK$), which resulted in densities of $\simeq80$ and $\simeq410$ atoms per $100\times100\unm^2$. We only considered pairs of adatoms in the analysis that retain a distance $\geq2.3\unm$ to other impurities. Within our resolution, this ensures the absence of any influence of other adatoms on the \YSR states of the dimers.  The differential conductance \didv was recorded using a standard lock-in technique with a frequency of 912\,Hz and a bias modulation amplitude of 15\,$\mu$V$_{rms}$. Energy resolution beyond the Fermi-Dirac limit is achieved by covering the etched W-tips with a layer of Pb until the tip shows bulk-like superconductivity~\cite{RubyPb15}. In combination with elaborate grounding and RF filtering, this allows us to reach an effective energy resolution of $\approx80\umueV$. In first approximation, measurements with a superconducting tip probe a convolution of tip and sample density of states, which shifts all spectral features by the tip's gap parameter $\pm\Delta_{\rm tip}/e$.


\begin{figure*}[]
	\includegraphics[width=0.98\textwidth]{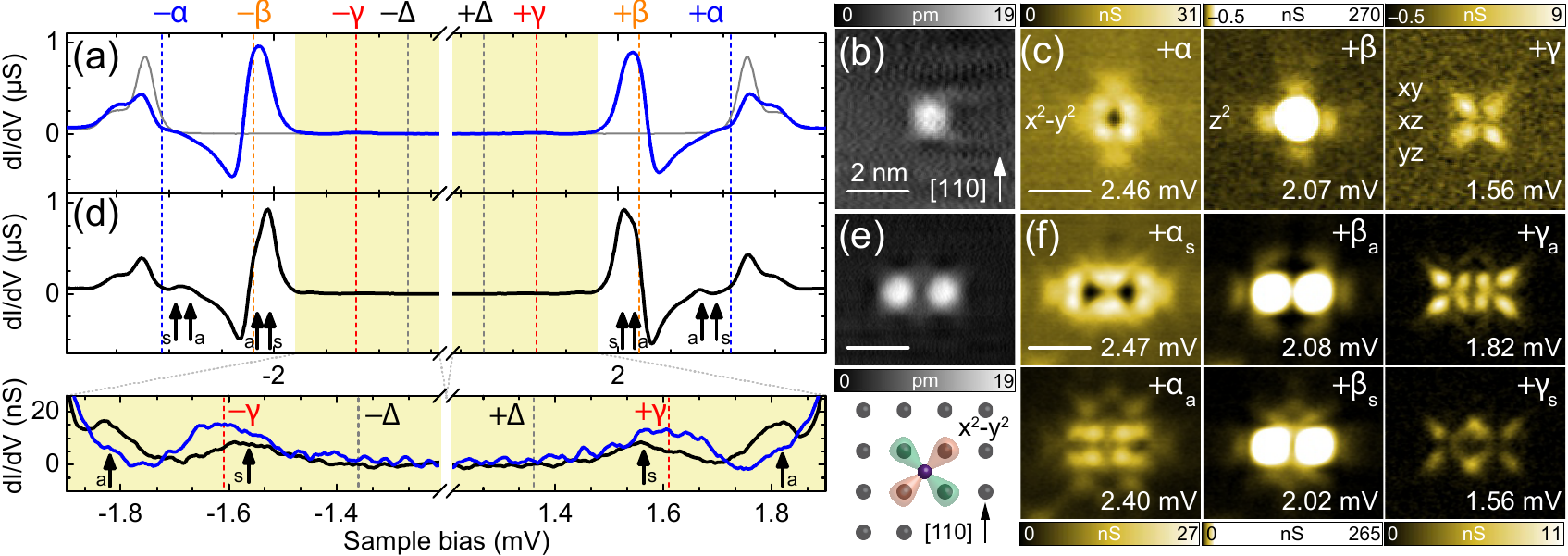}
	\caption{(a) \didv spectrum at the center of the single Mn adatom shown in (b). Three subgap resonances ($\pm\alpha$, $\pm\beta$, $\pm\gamma$) and the tip gap ($\pm\Delta$) are marked in the spectrum by dashed vertical lines (blue, orange, red, gray). For reference, a trace taken on the pristine substrate is superimposed (solid gray line). 
Set point: $300\upA$, $5\umV$. (c) \didv maps of a monomer at $+\alpha$, $+\beta$, and $+\gamma$, covering the same area as in (b). 
	(d) \didv spectrum of the dimer shown in (e), which is oriented along the \dir{\mbox{1$\bar10$}} direction and separated by $1.38 \pm 0.08\unm$. Each subgap state is split into two resonances $\alpha_\mathrm{s,a}$, $\beta_\mathrm{s,a}$, and $\gamma_\mathrm{s,a}$ (marked by arrows). Set point: $200\upA, 4\umV$.  (f) \didv maps taken at the positive-energy YSR resonances as marked in the figure. The scale for the resonances $\beta$ in (c) and for $\beta_\mathrm{a,b}$ in (f) is cut to emphasize the laterally extended intensity around the high intensity at the impurity center.}
	\label{fig:FigMonomerDimer}
\end{figure*}

We begin by reviewing the \YSR states of isolated Mn adatoms on Pb(001) (see \Fig{FigMonomerDimer}a) \cite{RubyMn16}. Differential conductance spectra acquired with a superconducting tip show two pairs of Bardeen-Cooper-Schrieffer (BCS) singularities near a sample bias of $\pm2.65\umV$ \cite{RubyPb15}. For a single Mn adatom, we find three additional pairs of \YSR resonances inside the superconducting gap (\Fig{FigMonomerDimer}a). Assuming that the Mn adatom is in a $d^5$ configuration, it conserves the orbital angular momentum of electrons in the superconductor and binds them in the $d$ channel \cite{Schrieffer67,RubyMn16}. The hollow adsorption site imposes a square pyramidal crystal field, which lifts the degeneracy of the $d$ states. Simple considerations of crystal field theory can be applied to deduce the order of the energy levels. The $d_{x^2-y^2}$ state lies highest in energy, followed by the $d_{z^2}$ orbital at an intermediate and the degenerate $d_{xy}$, $d_{xz}$, $d_{yz}$ states at the lowest energy.  This explains the characteristic shapes of the \YSR states in the \didv maps (\Figure{FigMonomerDimer}c) \cite{RubyMn16}. Moreover, the observation of distinct $d$-orbital-like bound-state patterns implies that Hund's energy is larger than the energy splitting of the adatom $d$ levels. The most intense resonance labeled by $\beta$ arises from the $d_{z^2}$ state. The faint resonance close to the superconducting gap edge (labeled $\alpha$) derives from the $d_{x^2-y^2}$ state, and the lowest lying resonance (labeled $\gamma$) is a mixture of scattering at the degenerate $d_{xy}$, $d_{xz}$, and $d_{yz}$ states. Tunneling into the $d_{xy}$ state is less favorable than into the $d_{xz}$, and $d_{yz}$ states, so that the \didv maps are dominated by the shapes of the $d_{xz}$- and $d_{yz}$-like orbitals  \cite{RubyMn16}. 

Now consider a dimer of Mn adatoms oriented along the \dir{\mbox{1$\bar10$}} direction (topography in \Fig{FigMonomerDimer}e). The Mn--Mn distance of $1.38 \pm 0.08\unm$ corresponds to a separation of four lattice spacings (i.e., the distance between nearest-neighbor adsorption sites along \udir{110}). \didv spectra on the adatoms of the dimer reveal that each single-atom \YSR resonance splits into two (\Fig{FigMonomerDimer}d). Moreover, \didv maps at the energies of the \YSR resonances again exhibit characteristic shapes (\Fig{FigMonomerDimer}f). 
Many features of the maps for individual atoms can be recognized. For instance, the clover shapes of the $d_{x^2-y^2}$ and of the $d_{xz}$, $d_{yz}$ states are still seen in the split $\alpha$ and $\gamma$ states, respectively. The strong intensity of the $d_{z^2}$-derived \YSR resonance $\beta$ is also found on the dimer constituents. 

However, a more detailed inspection reveals distinct differences between the maps for monomers and dimers. 
This is most clearly observed for the split $\gamma$ resonance. The resonance $+\gamma_{s}$ exhibits two pairs of overlapping lobes in between the adatoms which are increased in intensity compared to the outer lobes. In contrast, $+\gamma_{a}$ has outer lobes of increased intensity, while the inner lobes have reduced intensity and do not overlap. There is an apparent nodal line perpendicular to the dimer axis. Similar behavior is also observed for $\alpha$, where the overall intensity is shifted outwards for $\alpha_a$, but inwards for $\alpha_s$. Only minor variations are observed for $\beta$, yet with a similar trend and a nodal line in the case of $\beta_a$. We interpret these modified intensity distributions as fingerprints of  symmetric ($s$) and antisymmetric ($a$) combinations of YSR wave functions, hence the indices used above. Interestingly, while the antisymmetric $+\gamma_{a}$ and $+\beta_{a}$ resonances have higher energy than $+\gamma_{s}$ and  $+\beta_{s}$,respectively, it is the symmetric state that is higher in energy in the case of  $+\alpha$. The relatively small energy splittings and the preservation of the characteristic orbital shapes indicate a small hybridization strength which does not lead to a change in the order or a mixture of \YSR states derived from the individual adatoms.  

\begin{figure*}[t]
	\includegraphics[width=0.98\textwidth]{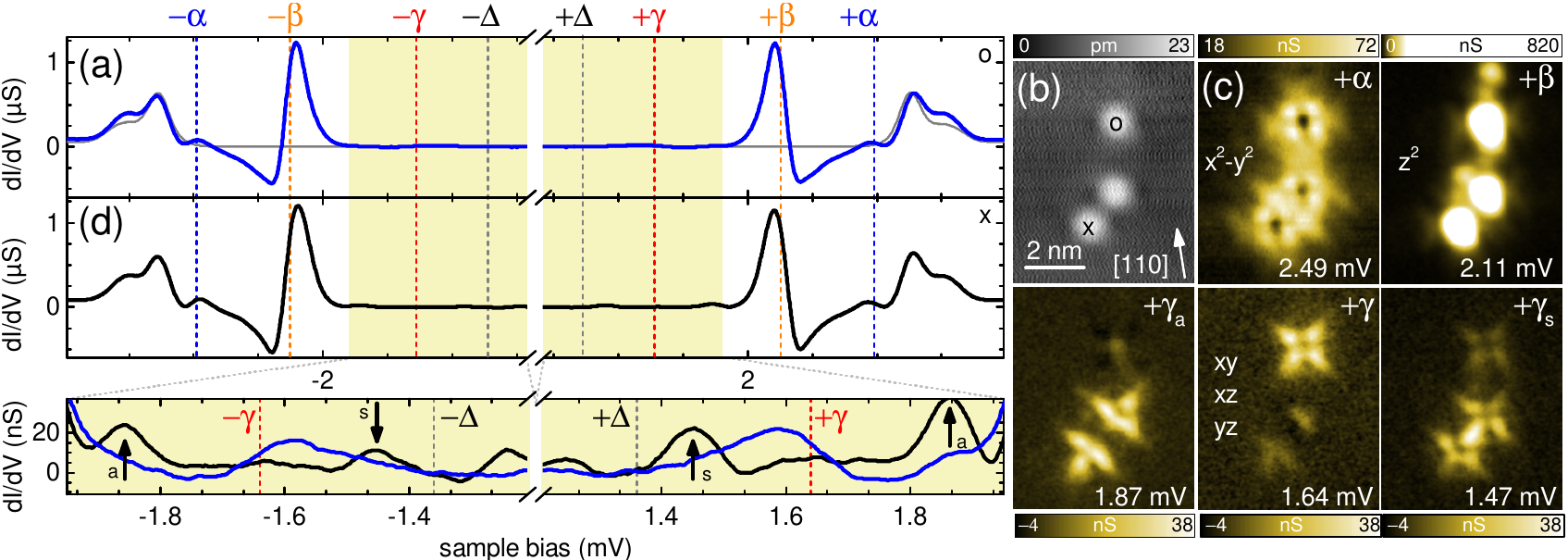}
	\caption{(a) \didv spectrum at the center of the single Mn. Three subgap resonances ($\pm\alpha$, $\pm\beta$, $\pm\gamma$) and the tip gap ($\pm\Delta$) are marked in the spectrum by dashed vertical lines (blue, orange, red, gray). For reference, a trace taken on the pristine substrate is superimposed (solid gray line). 
Set point: $300\upA$, $5\umV$. (b) Topography and (c) \didv maps of three Mn adatoms, two of which form a dimer oriented along the [100] direction and separated by $1.47 \pm 0.08\unm$. 
	(d) \didv spectrum of the dimer shown in (b). $\gamma$ is split into two resonances $\gamma_\mathrm{s,a}$ (marked by arrows). A faint signal at the energy of the monomer $\pm \gamma$ peaks hints at a third resonance. Set point: $200\upA, 4\umV$. }
	\label{fig:FigMonomerDimer100}
\end{figure*}

These observations suggest that to a good approximation, we can describe the coupled \YSR states as linear combinations of individual \YSR wave functions. Moreover, a splitting of \YSR states can only occur if their spin wave functions are not orthogonal \cite{Yao14, Hoffman15}. This implies that the alignment of the adatom spins is not antiferromagnetic, consistent with theoretical expectations \cite{Yao14}. The energy of the molecular \YSR states can be obtained by analogy to the linear combination of atomic orbitals in an H$_2$ molecule (for details, see Supplemental Material \cite{Supplementary}). This yields $E_\pm=E_s +\frac{C\pm D}{1+S}$ for  the energies of the \YSR states. Here, $E_s$ denotes the energy of the single impurity, $C_{ij}=\int d\boldsymbol{r}\,J(\boldsymbol{r}+d{\bf \hat{y}})\phi_{i}^\dagger(\boldsymbol{r})\phi_j(\mathbf{r})$ a Coulomb-like integral, $D_{ij}=\int d\boldsymbol{r}\,J(\boldsymbol{r})\phi_i(\boldsymbol{r})^{\dagger}\phi_j(\boldsymbol{r}+d{\bf \hat{y}})$ an exchange-like integral, and $S_{ij}=\int d\boldsymbol{r}\,\phi_{i}(\boldsymbol{r})^{\dagger}\phi_{j}(\boldsymbol{r}+d{\bf \hat{y}})$ an overlap integral with $\phi_{i,j}$ being the \YSR wave function deriving from one of the five $d$ orbitals. We notice that the Coulomb-like integral $C$ provides a  shift and the exchange-like integral $D$ produces a splitting. $C$ falls off monotonously with distance  $d{\bf \hat{y}}$ (choosing the dimer axis along the $y$ direction) and has the same sign as $J(\boldsymbol{r})$. It is thus positive or negative depending on whether the coupling between the impurity and the itinerant electrons is antiferromagnetic or ferromagnetic. The sign of the exchange-like integral $D$ alternates as a function of separation $d$ because the \YSR wave function $\phi(\boldsymbol{r})$ oscillates with the Fermi wavelength $\lambda_F$.  Hence, unlike the case of atomic orbitals in H$_2$, the order in energy does not reflect whether the wave function is symmetric (without a nodal plane) or antisymmetric (with a nodal plane).  In view of our experimental results, this explains why $+\gamma_a$ and $+\beta_a$ have a larger energy than  $+\gamma_s$ and $+\beta_s$, respectively, whereas the order of symmetric and antisymmetric \YSR wave functions is reversed in the case of $\alpha$ with $+\alpha_s$ having larger energy than $+\alpha_a$. One should therefore avoid calling these states bonding or antibonding.

To further validate these interpretations, we also investigated Mn dimers which are oriented along the \udir{100} directions of the Pb lattice. \Figure{FigMonomerDimer100} shows experimental results for such a dimer with a separation of  $1.47\pm0.08$\,nm along the $y$-axis or three lattice spacings along \udir{100}. The \didv spectrum in \Fig{FigMonomerDimer100}d shows no splitting for the $\alpha$ and $\beta$ resonances, and the corresponding \didv maps in \Fig{FigMonomerDimer100}c resemble simple superpositions of the single-adatom maps. (Note also that the third adatom in the vicinity exhibits the spectrum of an isolated Mn adatom.) Unlike $\alpha$ and $\beta$, the $\gamma$ resonance shows a sizable splitting into two as well as hints of an additional resonance which remains unshifted relative to the $\gamma$ resonance of the monomer. The absence of a hybridization shift suggests that the latter resonance could be associated with the $d_{xz}$-like \YSR state which is expected to have the smallest overlap. The corresponding \didv map shows faint intensity consistent with the shape of the $d_{xz}$-like \YSR state (\Fig{FigMonomerDimer100}c bottom, middle). The split-off $\gamma$ resonances would then originate from linear combinations of the $d_{yz}$- and $d_{xy}$-like \YSR states which have hybridizations of (nearly) equal strength. This is consistent with the strong intensity along the bonding direction of the split-off resonance deeper inside the superconducting gap ($\gamma_s$) indicating a symmetric combination of monomer states. Similarly, the \didv map of the resonance closer to the gap edge ($\gamma_a$) is reminiscent of the antisymmetric combination. The observed intensity perpendicular to the bonding direction would then originate from the $d_{xy}$-like \YSR states, possibly distorted by the Pb atom lying on the dimer axis. This interpretation is in agreement with theoretical symmetry considerations (see Supplemental Material).

\begin{figure}[b]
	\includegraphics[width=0.48\textwidth]{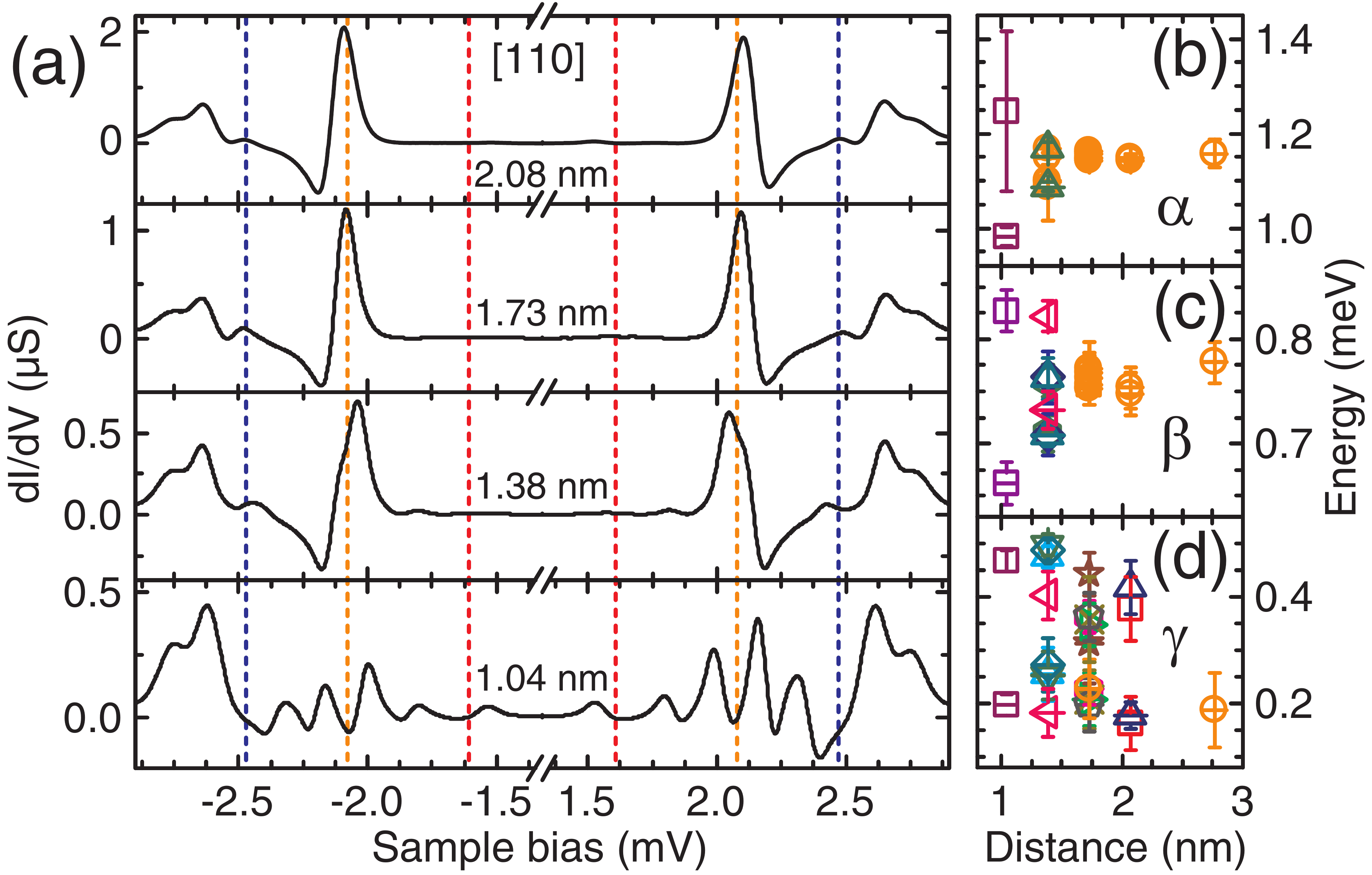}
	\caption{(a) \didv spectra of  Mn dimers with different interatomic distances oriented along the \udir{110} direction. 
Spectra are recorded at the center of one of the adatoms of a pair. Setpoint: $200\upA, 4\umV$. (b -- d) provide the splitting in energy of the peaks $\alpha$, $\beta$ and $\gamma$ as a function of the interatomic distance for the  \udir{110} direction. Same color and symbol indicate split pairs of resonances from the same dimer spectrum. Different symbols/colors correspond to data from different dimers except for yellow circles, which indicate data points where no splitting was observed for the respective resonance. }
	\label{fig:110Distance}
\end{figure}

\begin{figure}[tb]
		\includegraphics[width=0.48\textwidth]{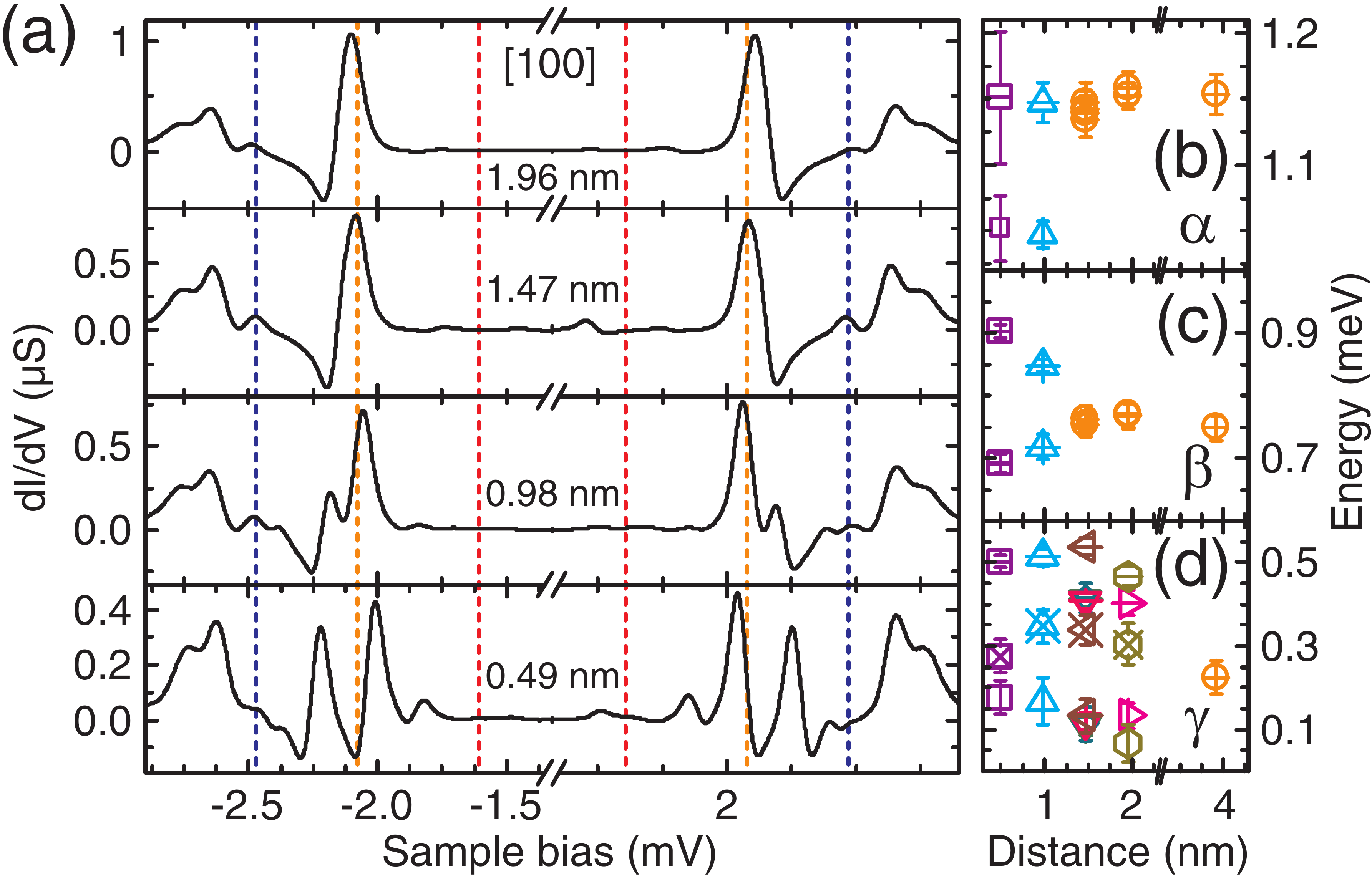}
	\caption{(a) \didv spectra of  Mn dimers with different interatomic distances oriented along the \udir{100} direction. 
Spectra are recorded at the center of one of the adatoms of a pair. Setpoint: $200\upA, 4\umV$. (b - d) give the splitting in energy of the peaks $\alpha$, $\beta$ and $\gamma$ as a function of the interatomic distance for this orientation. Same color code as in Fig.\ref{fig:110Distance} (b-d).  }
	\label{fig:100Distance}
\end{figure}

Figures \ref{fig:110Distance} and \ref{fig:100Distance} collect experimental results for the separation dependence of the resonance splittings. \Fig{110Distance} focuses on dimers oriented along \udir{110}. Panel (a) shows four representative spectra for separations corresponding to three to six lattice spacings along \udir{110}. Panels (b)-(d) collect the \YSR resonance energies for additional dimers. For adatom separations of $d=2.77\unm$ (eight lattice spacings), none of the \YSR resonances is split within our energy resolution of $\approx 80\umueV$. The splitting of the $d_{x^2-y^2}$-derived \YSR resonance $\alpha$ is resolved for one (out of four) of the observed dimers with a separation of $1.38\unm$ (four lattice spacings). For smaller adatom distances, we resolve the splitting in all dimers, with splittings of $\approx 0.2\umeV$ for $d=1.04\unm$ (three lattice spacings). The splitting of the $d_{z^2}$-derived \YSR resonance sets in at the same separation and is of approximately the same magnitude. In comparison, the  $d_{xy}$, $d_{xz}$, $d_{yz}$-derived \YSR resonance $\gamma$ already splits at larger distances ($d<2.08\unm$), with splittings up to $\approx 0.3\umeV$ for the smallest dimers.

The splittings of the \YSR resonances in dimers oriented along \udir{100} show similar behavior. Figure \ref{fig:100Distance} shows four representative spectra as well as the extracted energy positions of the \YSR resonances. The splitting of the $d_{x^2-y^2}$- and the $d_{z^2}$-derived \YSR resonance is only observable for $d\leq 0.98\unm$ (two lattice spacings) with a splitting of $\approx 0.1$ and $0.2$\,meV, respectively, at $d=0.49$\,nm (one lattice spacing). The extracted $d_{x^2-y^2}$-derived resonances hint at an overall downward shift with decreasing distance. As already described above, we observe a splitting of the $\gamma$ resonance into three components for many (though not all) dimers with the central resonance remaining at the energy of the monomer (see the discussion of the faint resonances at $\pm \gamma$ seen in \Fig{FigMonomerDimer100}).

In addition to the decay with adatom separation, theory predicts oscillatory behavior of the energy splitting with a period of half the Fermi wave length $\lambda_F$ (see discussion above and Supplemental Material). For Pb, $\lambda_F/2$ of the outer Fermi sheet, which gives rise to the \YSR states~\cite{RubyMn16}, equals $0.61\pm0.03 \unm$ along the \udir{110} direction and $0.53\pm0.06 \unm$ along the \udir{100} direction~\cite{Lykken1971}. The range over which we resolve the energy splitting is only slightly larger than $\lambda_F/2$ and contains only three distinct separations due to the discreteness of the adsorption sites. This precludes testing the oscillatory behavior of the \YSR splitting in our experiments. Moreover, we may only extract a hint of a distance-dependent shift of the center of mass of the \YSR resonances for the $\gamma$ resonances of the \udir{110} dimers. Depending on the particular resonance, theory predicts a shift of at most one quarter of the energy splitting (see Supplementary Information \cite{Supplementary}), which is at the limit of our energy resolution.

In conclusion, we resolved and analyzed the hybridization of \YSR states originating from Mn adatoms which are located three to six lattice spacings apart on Pb(001) and observe characteristic energy splittings of up to a few hundred $\umueV$. At these relatively large distances, direct exchange coupling or simple superexchange via a single substrate atom can be neglected. Instead, we show by mapping the spatial distribution of the dimer \YSR states that the coupling hybridizes monomer \YSR states into symmetric and antisymmetric linear combinations. The observed hybridization precludes antiferromagnetic alignment of the adatom magnetic moments. We have also recorded \didv spectra with a spin-polarized tip, but did not observe any spin contrast with oppositely magnetized tips or varying contrast in different dimers. This suggests that the spin orientation fluctuates due to thermal excitations. 

The hybridization strength is comparable to the RKKY coupling on normal metal surfaces~\cite{Zhou10}. When coupling not only two adatoms but rather an entire chain, one expects the formation of \YSR bands. These may give rise to topological superconductivity and an alternative route towards the realization of Majorana bound states ~\cite{NadjPerge2013, Pientka2013, Schechter2016}. To date, adatom-based Majorana experiments rely on compact ferromagnetic chains, in which the direct coupling of adatom $d$-states is presumably essential for the formation of a topological superconducting phase~\cite{NadjPerge14, RubyMaj15, Pawlak16, Feldman16,Ruby17, Jeon2017}.


We gratefully acknowledge funding by the Deutsche Forschungsgemeinschaft through HE7368/2, FR2726/4, and CRC 183, as well as the ERC consolidator grant NanoSpin.

\bibliographystyle{apsrev4-1}

\clearpage

\setcounter{figure}{0}
\setcounter{section}{0}
\setcounter{equation}{0}
\renewcommand{\theequation}{S\arabic{equation}}
\renewcommand{\thefigure}{S\arabic{figure}}

\onecolumngrid

\renewcommand{\Fig}[1]{\mbox{Fig.\unitspace\ref{Sfig:#1}}}
\renewcommand{\Figure}[1]{\mbox{Figure\unitspace\ref{Sfig:#1}}}

\section*{\Large{Supplemental Material}}

\vspace{0.7cm}
\section{Shiba state with a single magnetic impurity}
\subsection{General Consideration}

To generate $d$-orbital-like Shiba bound states numerically (without attempting to accurately describe the specific system at hand), consider a single magnetic moment embedded in a homogeneous $s$-wave superconductor, as described by the Bogoliubov--de Gennes Hamiltonian
\begin{equation}
	H = H_{s} + J(r)\sigma_z,
\end{equation}
where $J(r)$ is the exchange potential between the magnetic moment and the itinerant electrons of the superconductor. We choose $J(r)$ isotropic and neglect the potential scattering by the impurity for simpliciity.  Note that we choose the direction of the magnetic moment to be the $z$ direction. The superconductor is described by the Hamiltonian
\begin{equation}
	H_s = \left(-\frac{\nabla^2}{2} - \mu\right)\tau_z + \Delta\tau_x.
\end{equation}
Here, $\sigma_{x,y,z}$ and $\tau_{x,y,z}$ are Pauli matrices in spin and particle-hole space, respectively, 
$\Delta>0$ is the pairing potential and $\mu$ the chemical potential. We choose units such that the electron charge $e$, the electron mass $m$, and $\hbar$ are all equal to unity.

Since the exchange potential $J(r)$ is isotropic, we use spherical coordinates $(r,\theta,\phi)$ centered at the position of the magnetic moment. The Hamiltonian for the superconductor $H_s$ can be rewritten as
\begin{equation}
	H_s = H_r\tau_z + \Delta \tau_x ,
\end{equation}
with
\begin{equation}
	H_{r} = -\frac{1}{2}\left( \frac{d^2}{dr^2} + \frac{2}{r}\frac{d}{dr} + \frac{l(l+1)}{r^2}\right) - \mu.
\end{equation}
where $l$ denotes the angular momentum.

Confining the system to a large sphere with radius $R$, one has a discrete set of basis functions
$\{\rho_{k,l}(r)Y_{l,m}(\theta,\phi)\}$, with spherical Harmonics $Y_{l,m}(\theta,\phi)$ and
\begin{equation}
	\rho_{k,l}(r) = \frac{\sqrt{2}}{\sqrt{R^3}}j_l(\alpha_{k,l}\frac{r}{R})/j_{l+1}(\alpha_{k,l}) 
	= \frac{\sqrt{2}}{\sqrt{R}r}J_{l+\frac{1}{2}}(\alpha_{k,l}\frac{r}{R})/J_{l+\frac{3}{2}}(\alpha_{k,l}).
	\label{eq:rhokl}
\end{equation}
Here, $j_{l}$ and $J_l$ are the spherical and cylindrical Bessel function of order $l$. $j_{l}(\alpha_{k,l})$
is normalized in the sphere of radius $R$ and $\alpha_{k,l}$ is the $k$th zero of $j_l$. We have used the relation
\begin{equation}
	j_{l}(r) =  \sqrt{\frac{\pi}{2r}}J_{l+\frac{1}{2}}(r)
\end{equation}
in obtaining the above equation.

Since the Hamiltonian is isotropic, it is block-diagonal in the angular-momentum quantum numbers $l,m$. 
For each $l,m$, the Hamiltonian $H_s$ of the superconductor is diagonal in $k$ with matrix elements
\begin{equation}
	(H_s)_{k,k'} = \left[(\frac{\alpha_{k,l}^2}{2R^2} - \mu)\tau_z + \Delta\tau_x\right]\delta_{k,k'}.
\end{equation}
The exchange potential has matrix elements 
\begin{equation}
	J_{k,k'} = \int_{0}^{\infty} r^2 dr \rho_{k,l}(r)\rho_{k',l}(r)J(r).
\end{equation}
To find the Shiba state, we fix $\sigma_z = 1$ and solve the eigenvalue problem with eigenvalue $-\Delta<E<\Delta$. The other solution at the opposite energy follows from $\sigma_z=-1$ and can be obtained by particle-hole symmetry.

\subsection{Shiba states with $l=2$}

To simulate the Shiba states of Mn adatoms, we consider the $l=2$ channel. For an adatom located in a completely isotropic environment, there are five degenerate Shiba states with the same radial wavefunction but different angular wavefunctions corresponding to $m=\pm 2, \pm 1, 0$. Instead of complex spherical harmonics, we can pass to the real angular-momentum basis, with
\begin{align}
  Y_{xy} &= \frac{i}{\sqrt{2}}\left( Y_{2,-2} - Y_{2,2} \right) \\
  Y_{yz} &= \frac{i}{\sqrt{2}}\left( Y_{2,-1} + Y_{2,1} \right) \\
  Y_{z^2} &= Y_{2,0} \\
  Y_{xz} &= \frac{1}{\sqrt{2}} \left( Y_{2,-1} - Y_{2,1} \right) \\
  Y_{x^2-y^2} &= \frac{1}{\sqrt{2}}\left( Y_{2,-2} + Y_{2,2}  \right).
  \label{eq:real_angular}
\end{align}
If we choose the quantization axis along the $z$-axis, these five wavefunctions have the shape of
$d_{xy}$, $d_{yz}$, $d_{z^2}$, $d_{xz}$ and $d_{x^2-y^2}$ orbitals, respectively. 

In experiment, the Mn adatom is located on the surface of a superconductor, which reduces the symmetry of the adatom environment to the point group $C_{4v}$. Thus, the five degenerate Shiba states split
due to the crystal field according to the irreproducible representations of $C_{4v}$.  
If we take the $z$-direction along the normal to the surface of the superconductor, the $d_{z^2}$, $d_{x^2-y^2}$, and $d_{xy}$ states are nondegenerate, while the $d_{xz}$ and $d_{yz}$ are degenerate. Experiment yields only three peaks as the $d_{xz}$, $d_{yz}$, and $d_{xy}$ states are close in energy [see Ref.\cite{Ruby2016}]

\section{Shiba state with two magnetic impurities}

\subsection{Variational ansatz for Shiba dimer wavefunction}

Now consider a system with two ferromagnetically aligned magnetic impurities embedded in a superconductor. Motivated by our experimental results, we assume that the coupling between the two adatoms is weak compared to the energy separation between the $\alpha$, $\beta$, and $\gamma$ peaks. 
In this limit, the wavefunctions of magnetic dimers can be written as linear combinations
of Shiba states of the individual impurities. 

For two magnetic impurities, the Hamiltonian can be written as
\begin{equation}
H=H_{s}+J(\boldsymbol{r}-\frac{d}{2}\hat{y})\sigma_{z}+J(\boldsymbol{r}+\frac{d}{2}\hat{y})\sigma_{z},
\label{eq:dimer_Hamiltonian}
\end{equation}
where $d$ denotes the distance between the two impurities. We choose the dimer axis to be aligned along the $y$ axis. Similar to the discussion for a single impurity, we can fix $\sigma_{z}=1$. 

When the two magnetic impurities couple, the single-impurity peaks in the STM measurement split 
due to hybridization of the corresponding single Shiba wavefunctions. Hence, we make the variational ansatz 
\begin{equation}
	\psi(r)=\sum_{j}\left\{c_{1,j}\phi_j(\boldsymbol{r}-\frac{d}{2}\hat{y})+c_{2,j}\phi_j(\boldsymbol{r}+\frac{d}{2}\hat{y})\right\},
\end{equation}
 for the dimer wavefunction. Here, $\phi_j(\boldsymbol{r})$ is the two component Shiba wave function for a single impurity with $j=z^2,x^2-y^2,xy,yz,xz$, which satisfies
\begin{equation}
	\left[H_{s}+J(\boldsymbol{r})\right]\phi_j(\boldsymbol{r})=E_{s}\phi_j(\boldsymbol{r}), \quad \abs{E_s}\leq \Delta.
\end{equation}
The sum over $j$ refers to the sum over $xy,yz,xz$ for the $\gamma$ peak, and involves only the $x^2-y^2$ and $z^2$ orbitals for the $\alpha$ and $\beta$ peaks, respectively. The Shiba energy $E_s$ for a single impurity can be obtained numerically following the discussion in the previous section.

Using the variational wave function, we obtain the following generalized eigenvalue equation
\begin{equation}
\left(\begin{array}{cc}
	E_{s}\mathbf{1}+\mathbf{C} & E_{s}\mathbf{S}+\mathbf{D}\\
	E_{s}\mathbf{S}+\mathbf{D} & E_{s}\mathbf{1}+\mathbf{C}
\end{array}\right)\left(\begin{array}{c}
	\mathbf{c}_{1}\\
	\mathbf{c}_{2}
\end{array}\right)=
\left(\begin{array}{cc}
	E\mathbf{1}& E\mathbf{S}\\
	E\mathbf{S}& E\mathbf{1}
\end{array}\right)
\left(\begin{array}{c}
	\mathbf{c}_{1}\\
	\mathbf{c}_{2}
\end{array}\right)
\label{eq:g_eigen}
\end{equation}
where $\mathbf{S}$, $\mathbf{C}$ and $\mathbf{D}$ are matrices for overlap, Coulomb-like and exchange-like integrals, similar to the integrals describing the chemical bonding of the $\mathrm{H}_2$ molecule. The corresponding matrix elements are given by 
\begin{gather}
	S_{ij}=\int d\boldsymbol{r}\,\phi_{i}(\boldsymbol{r})^{\dagger}\phi_{j}(\boldsymbol{r}+d\hat{y})\\
	C_{ij}=\int d\boldsymbol{r}\,J(\boldsymbol{r}+d\hat{y})\phi_{i}^\dagger(\boldsymbol{r})\phi_j(\mathbf{r})\\
	D_{ij}=\int d\boldsymbol{r}\,J(\boldsymbol{r})\phi_i(\boldsymbol{r})^{\dagger}\phi_j(\boldsymbol{r}+d\hat{y}).
\end{gather}
$\mathbf{c}_{1}$ and $\mathbf{c}_{2}$ are column vectors with elements $c_{1,j}$ and $c_{2,j}$ respectively, in which $j$ takes values from the relevant subset of the five $d$ states, depending on the degeneracy. 

For the nondegerate $\alpha$ and $\beta$ peaks, the above matrices and vectors are only scalars. We denote theses scalars without their indices for simplicity. By solving the eigenvalue equation, we obtain two Shiba energies
\begin{equation}
	E_{\pm}= E_{s}+\frac{C\pm D}{1\pm S}
	\label{eq:Epm_full}
\end{equation}
with wavefunctions
\begin{equation}
\psi=\phi(\boldsymbol{r}-\frac{d}{2}\hat{y})\pm \phi(\boldsymbol{r}+\frac{d}{2}\hat{y}).
\end{equation}

When the separation $d$ is large, one has $S$$\ll1$ and obtains two
bound states with energies 
\begin{equation}
E_{\pm}=E_{s}+C\pm D.
\label{eq:Epm_appr}
\end{equation}

\subsection{Evaluating integrals}

To obtain the matrices $\mathbf{S}$, $\mathbf{C}$, and $\mathbf{D}$, we need to evaluate integrals which involve functions centered at two locations with separation $d$. We first consider the simple case, in which the two impurities are aligned along $z$ instead of the $y$ axis. The result for the latter case can then be obtained via Wigner rotations.

Imagine we have two coordinate systems $a$ and $b$ for which the $z$-axis coincides with the dimer axis, the $x,y$ axes of the two coordinate systems are parallel to each other, and the origins coincide with the adatom locations. In terms of spherical coordinates, a point in space can be written as $(r_a,\theta_a,\phi)$ and $(r_b,\theta_b,\phi)$. It is convenient to introduce prolate spheroidal coordinates $(\xi,\eta,\phi)$, defined by
\begin{equation}
r_{a}=\frac{\xi+\eta}{2}d,\quad r_{b}=\frac{\xi-\eta}{2}d,\quad\cos\theta_{a}=\frac{\xi\eta+1}{\xi+\eta},\quad\cos\theta_{b}=\frac{\xi\eta-1}{\xi-\eta},\quad
dV=\frac{d^{3}}{8}(\xi^{2}-\eta^{2})d\xi d\eta d\phi,
\end{equation}
where $dV$ is the volume element.

Denote the Shiba wave function for a single impurity as
\begin{equation}
\phi(\mathbf{r})=R(r)Y_{lm}(\hat{r}),
\end{equation}
where $R(r)$ is a two-component radial wavefunction and the complex spherical harmonics is defined as
\begin{equation}
  Y_{lm}(\hat{\mathbf{r}})=A_{lm}P_{l,m}(cos(\theta))e^{im\phi},\quad
  A_{lm}=\sqrt{\frac{2l+1}{4\pi}\frac{(l-m)!}{(l+m)!}},
\end{equation}
with $P_{l,m}(x)$ the associated Legendre polynomial.

In the following, we first evaluate the integrals with Shiba states as given above. For the case with the two impurities aligned along the $z$ axis, we denote the matrices as $\mathbf{S}'$, $\mathbf{C}'$, and $\mathbf{D}'$.  

\subsubsection{Overlap integral $\mathbf{S}'$}

\begin{gather}
S_{ab}'=\int
d\boldsymbol{r}\,\phi_{a}(\boldsymbol{r})^{\dagger}\phi_{b}(\boldsymbol{r}+d\hat{z})=A_{la,ma}A_{lb,mb}\frac{2\pi
  d^{3}}{8}\delta_{ma,mb}I_{S}(d) \label{eq:Sint}\\
I_{S}(d)=\int_{-1}^{1}d\eta\int_{1}^{\xi^{*}}d\xi\alpha_{S}(\eta,\xi;d)\\
\alpha_{S}(\eta,\xi;d)=R_{a}(\frac{(\xi+\eta)d}{2})^{\dagger}R_{b}(\frac{(\xi-\eta)d}{2})P_{la,ma}(\frac{1+\xi\eta}{\xi+\eta})
P_{lb,mb}(\frac{\xi\eta-1}{\xi-\eta})(\xi^{2}-\eta^{2})
\end{gather}

\subsubsection{Coulomb-like integral $\mathbf{C}'$}
\begin{gather}
C_{ab}'=\int
d\boldsymbol{r}\,\phi_{a}(\boldsymbol{r})^{\dagger}\phi_{b}(\boldsymbol{r})J(\boldsymbol{r}+d\hat{z})=A_{la,ma}A_{lbmb}\frac{2\pi
  d^{3}}{8}\delta_{ma,mb}I_{C}(d)\label{eq:Cint}\\
I_{C}(d)=\int_{-1}^{1}d\eta\int_{1}^{\infty}d\xi\alpha_{C}(\eta,\xi;d)\\
\alpha_{C}(\eta,\xi;d)=R_{a}(\frac{(\xi+\eta)d}{2})^{\dagger}R_{b}(\frac{(\xi+\eta)d}{2})P_{la,ma}(\frac{1+\xi\eta}{\xi+\eta})P_{lb,mb}(\frac{1+\xi\eta}{\xi+\eta})J(\frac{(\xi-\eta)d}{2})(\xi^{2}-\eta^{2})
\end{gather}

\subsubsection{Exchange-like integral $\mathbf{D}'$}
\begin{gather}
  D_{ab}'=\int
  d\boldsymbol{r}\,\phi_{a}(\boldsymbol{r})^{\dagger}\phi_{b}(\boldsymbol{r}+d\hat{z})J(\boldsymbol{r})=A_{la,ma}A_{lbmb}\frac{2\pi
    d^{3}}{8}\delta_{ma,mb}I_{D}(d)\label{eq:Dint}\\
I_{D}(d)=\int_{-1}^{1}d\eta\int_{1}^{\infty}d\xi\alpha_{D}(\eta,\xi;d)\\
\alpha_{D}(\eta,\xi;d)=R_{a}(\frac{(\xi+\eta)d}{2})^{\dagger}R_{b}(\frac{(\xi-\eta)d}{2})P_{la,ma}(\frac{1+\xi\eta}{\xi+\eta})
P_{lb,mb}(\frac{\xi\eta-1}{\xi-\eta})J(\frac{(\xi+\eta)d}{2})(\xi^{2}-\eta^{2}).
\end{gather}

We evaluate the two-dimensional integrals $I_{S}$, $I_{C}$ and $I_{D}$ numerically.

\subsubsection{Basis Transformation}

So far, we chose the angular-momentum quantization axis for a single-impurity Shiba state parallel to the dimer axis. Let us denote the Cartesian axes of this coordinate system by $x'y'z'$. We now evaluate the integrals in the $xyz$ coordinate system in which the two impurities are aligned along the $y$-axis as introduced in the Hamiltonian (\ref{eq:dimer_Hamiltonian}).

The spherical harmonics $\ket{lm}_{x'y'z'}$ in coordinate system $x'y'z'$ is related to the ones $\ket{lm}_{xyz}$ in coordinate system $xyz$ by a rotation $R(\varphi,\theta,\psi)$, where $(\varphi,\theta,\psi)=(0,\pi/2,\pi/2)$ are Euler angles, namely
\begin{equation}
	\ket{lm}_{xyz} = R(\varphi,\theta,\psi)\ket{lm}_{x'y'z'}=\sum_{m'} D^{l}_{m'm}(\varphi,\theta,\psi)\ket{lm'}_{x'y'z'}.
\end{equation}
Here, we introduced the Wigner matrix
\begin{equation}
	\mathcal{D}_{m'm}^l(\varphi,\theta,\psi) = \bra{lm'}R(\varphi,\theta,\psi)\ket{lm},
\end{equation}
which has the property
\begin{equation}
	\mathcal{D}_{m'm}^{l}(\varphi,\theta,\psi)=e^{-i\varphi m'} \mathcal{D}_{m'm}^{l}(0,\theta,0) e^{-i\psi m}.
\end{equation}

In $xyz$ coordinate, the matrices for overlap, Coulomb-like and exchange-like integrals computed above transform into
\begin{gather}
	\mathbf{S}^{\rm complex} = \mathcal{D}^{\dagger}\mathbf{S}'\mathcal{D} \\
	\mathbf{C}^{\rm complex} = \mathcal{D}^{\dagger}\mathbf{C}'\mathcal{D} \\
	\mathbf{D}^{\rm complex} = \mathcal{D}^{\dagger}\mathbf{D}'\mathcal{D},
\end{gather}
where $\mathbf{S}'$, $\mathbf{C}'$, and $\mathbf{D}'$ are the diagonal matrices given in 
Eqs.\, (\ref{eq:Sint}, \ref{eq:Cint}, and \ref{eq:Dint}), and the matrix $\mathcal{D}$ has matrix elements
\begin{equation}
	(\mathcal{D})_{m'm} = \mathcal{D}^{2}_{m'm}(\varphi,\theta,\psi),
\end{equation}
where $\alpha, \beta,\gamma$ are the Euler angles, which rotate $x'\to x$, $y'\to y$ and $z' \to z$.

Furthermore, we are interested in these matrices for the real angular basis as defined in Eq.\, (\ref{eq:real_angular}). In this real basis, we have
\begin{gather}
  \mathbf{S} = \mathcal{U}^{\dagger}\mathbf{S}^{\mathrm{complex}}\mathcal{U}=(\mathcal{DU})^\dagger\mathbf{S}'\mathcal{DU} \\
  \mathbf{C} = \mathcal{U}^{\dagger}\mathbf{C}^{\mathrm{complex}}\mathcal{U}=(\mathcal{DU})^\dagger\mathbf{C}'\mathcal{DU} \\
  \mathbf{D} = \mathcal{U}^{\dagger}\mathbf{D}^{\mathrm{complex}}\mathcal{U}=(\mathcal{DU})^\dagger\mathbf{D}'\mathcal{DU}, 
\end{gather}
where the matrix $\mathcal{U}$ is given by
\begin{equation}
  \mathcal{U}=\left(\begin{array}{ccccc}
\frac{i}{\sqrt{2}} & 0 & 0 & 0 & \frac{1}{\sqrt{2}}\\
0 & \frac{i}{\sqrt{2}} & 0 & \frac{1}{\sqrt{2}} & 0\\
0 & 0 & 1 & 0 & 0\\
0 & \frac{i}{\sqrt{2}} & 0 & -\frac{1}{\sqrt{2}} & 0\\
-\frac{i}{\sqrt{2}} & 0 & 0 & 0 & \frac{1}{\sqrt{2}}
\end{array}\right),
\end{equation}
and the real and complex angular bases are $\{Y_{xy},Y_{yz},Y_{z^2},Y_{xz}, Y_{x^2-y^2}\}$
and $\{Y_{2,-2},Y_{2,-1},Y_{2,0},Y_{2,1},Y_{2,2}\}$, respectively.

\subsection{Dimers aligned along $\braket{110}$ direction}

Now we consider the case where the dimers are aligned along the $\braket{110}$ direction. We need to take the Euler angles
$(\varphi,\theta,\psi)=(0,\pi/2,\pi/4)$, which gives rise to the rotation matrix
\begin{equation}
	\mathcal{D}=\left(\begin{matrix}\frac{i}{4} & \frac{e^{i\pi/4}}{2} & \frac{\sqrt{6}}{4} & -\frac{e^{i3\pi/4}}{2} & -\frac{i}{4}\\
-\frac{i}{2} & -\frac{e^{i\pi/4}}{2} & 0 & -\frac{e^{i3\pi/4}}{2} & -\frac{i}{2}\\
\frac{\sqrt{6}i}{4} & 0 & -\frac{1}{2} & 0 & -\frac{\sqrt{6}i}{4}\\
-\frac{i}{2} & \frac{e^{i\pi/4}}{2} & 0 & \frac{e^{i3\pi/4}}{2} & -\frac{i}{2}\\
\frac{i}{4} & -\frac{e^{i\pi/4}}{2} & \frac{\sqrt{6}}{4} & \frac{e^{i3\pi/4}}{2} & -\frac{i}{4}
\end{matrix}\right).
\end{equation}
After some algebra, we obtain the overlap matrix
\begin{equation}
	\boldsymbol{S}=\left(\begin{matrix}\frac{3S_{0}}{4}+\frac{S_{2}}{4} & 0 & \frac{\sqrt{3}}{4}\left(S_{0}-S_{2}\right) & 0 & 0\\
0 & \frac{S_{1}}{2}+\frac{S_{2}}{2} & 0 & -\frac{S_{1}}{2}+\frac{S_{2}}{2} & 0\\
\frac{\sqrt{3}}{4}\left(S_{0}-S_{2}\right) & 0 & \frac{S_{0}}{4}+\frac{3S_{2}}{4} & 0 & 0\\
0 & -\frac{S_{1}}{2}+\frac{S_{2}}{2} & 0 & \frac{S_{1}}{2}+\frac{S_{2}}{2} & 0\\
0 & 0 & 0 & 0 & S_{1}
\end{matrix}\right).
\label{eq:Smat_110}
\end{equation}
Similar expressions for $\mathbf{C}$ and $\mathbf{D}$ also exist and are obtained by simply replacing $S$ by $C$ and $D$, respectively.

\subsubsection{$\alpha$ and $\beta$ peaks}

At large distances, we apply Eq.~(\ref{eq:Epm_appr}) and obtain
\begin{gather}
	E^{\alpha}_{\pm} = E_{s}^{x^2-y^2} + C^{\alpha}_1 \pm D^{\alpha}_{1} \\
	E^{\beta}_{\pm} = E_{s}^{z^2} + \frac{(C^\beta_0 \pm D^\beta_0) + 3(C^\beta_2 - D^\beta_2)}{4}.
\end{gather}

\subsubsection{$\gamma$ peak}

Although the $\gamma$ peak derives from the $d_{xy}$, $d_{yz}$, and $d_{xz}$ orbitals, we see from  Eq.~(\ref{eq:Smat_110}) that $d_{xy}$ decouples from the others. We consider the large-$d$ case and neglect $S$. Using Eq.~(\ref{eq:Epm_appr}), we obtain
\begin{equation}
	E^{\gamma}_{xy,\pm} = E_{s}^{xy,yz,xz} + \frac{3(C_0^\gamma \pm D_0^\gamma) + (C_2^\gamma \pm D_2^\gamma)}{4},
\end{equation}
and the eigenstates are symmetric and antisymmetric superposition of the $d_{xy}$ states centered at the two adatoms.

To solve for the remaining eigenstates, one can introduce the new basis $\{\ket{d_+}, \ket{d_-}\}$, with
\begin{gather}
\ket{d_{+}} = \frac{1}{\sqrt{2}}(\ket{d_{yz}} + \ket{d_{xz}}) \\
\ket{d_{-}} = \frac{1}{\sqrt{2}}(\ket{d_{yz}} - \ket{d_{xz}}).
\end{gather}
In this new basis, $\mathbf{S}$, $\mathbf{C}$ and $\mathbf{D}$ become diagonal when restricted to the subspace spanned by $\ket{d_{yz}}$ and $\ket{d_{xz}}$. Thus, the $\ket{d_{+}}$ and $\ket{d_{-}}$ states in one adatom couple independently to the same states in the other adatom. We then obtain the energies
\begin{equation}
  E^{\gamma}_{+,\pm} = E_{s}^{xy,yz,xz} + C_2^\gamma \pm D_2^\gamma
\end{equation}
of the bound states, which correspond to symmetric and antisymmetric superpositions of $d_{+}$ states centered at the two adatoms, and the energies
\begin{equation}
  E^{\gamma}_{-,\pm} = E_{s}^{xy,yz,xz} + C_1^\gamma \pm D_1^\gamma,
\end{equation}
for the eigenstates which are symmetric and antisymmetric superposition of $d_{-}$ states centered at the two adatoms.

\subsection{Dimers aligned along $y$-axis ($\braket{100}$)}

In this case, we have $(\varphi,\theta,\psi)=(0,\pi/2,\pi/2)$. Thus,
\begin{equation}
\mathcal{D}=\left(\begin{array}{ccccc}
1/4 & 1/2 & \sqrt{6}/4 & 1/2 & 1/4\\
-1/2 & -1/2 & 0 & 1/2 & 1/2\\
\sqrt{6}/4 & 0 & -1/2 & 0 & \sqrt{6}/4\\
-1/2 & 1/2 & 0 & -1/2 & 1/2\\
1/4 & -1/2 & \sqrt{6}/4 & -1/2 & 1/4
\end{array}\right),
\end{equation}
which gives rise to
\begin{equation}
	\boldsymbol{S}=\left(\begin{matrix}S_{1} & 0 & 0 & 0 & 0\\
0 & S_{1} & 0 & 0 & 0\\
0 & 0 & \frac{S_{0}+3S_{2}}{4} & 0 & \frac{\sqrt{3}(S_{0}-S_{2})}{4}\\
0 & 0 & 0 & S_{2} & 0\\
0 & 0 & \frac{\sqrt{3}(S_{0}-S_{2})}{4} & 0 & \frac{3S_{0}+S_{2}}{4}
\end{matrix}\right)
\label{eq:Smat}
\end{equation}
Here $S_i=S_{ii}'$ are integrals defined in Eq.~(\ref{eq:Sint}) using complex spherical harmonics in the $x'y'z'$ coordinate system. Similar expressions for $\mathbf{C}$ and $\mathbf{D}$ also exist, by replacing $S_i$ by $C_i=C_{ii}'$ and $D_i=D_{ii}'$, which are given in Eqs.~(\ref{eq:Cint},\ref{eq:Dint}). We also used the relations $S_{i} = S_{-i}$, $C_{i} = C_{-i}$ and $D_i=D_{-i}$.
Now, we are in a position to analyze the $\alpha$, $\beta$, and $\gamma$ peaks separately.

\subsubsection{$\alpha$ and $\beta$ peaks}

Since the $\alpha$ and $\beta$ peaks derive from $d_{x^2-y^2}$ and $d_{z^2}$, respectively, we use
Eq.~(\ref{eq:Epm_full}) to compute the bound-state energy of the dimer. The corresponding matrix elements are
\begin{gather}
	S_{x^2-y^2,x^2-y^2} = \frac{3S_0+S_2}{4} \\
	S_{z^2,z^2} = \frac{S_0+3S_2}{4}.
\end{gather}
Similar expressions also exist for $C$ and $D$. Note that the integrals $S_i$, $C_i$ and $D_i$ depend on the radial wave functions, which are different for the different states. 

At large distance, i.e., when the overlap integrals can be neglected, one can apply Eq.~(\ref{eq:Epm_appr}).
We have
\begin{gather}
	E^{\alpha}_{\pm} = E_{s}^{x^2-y^2} + \frac{3(C^{\alpha}_0\pm D^{\alpha}_{0})+(C^{\alpha}_2 \pm D^\alpha_2)}{4}, \\
	E^{\beta}_{\pm} = E_{s}^{z^2} + \frac{(C^{\beta}_0\pm D^{\beta}_{0})+ 3(C^\beta_2 \pm D^\beta_2)}{4},
\end{gather}
where the subscripts $\alpha,\beta$ were added to distinguish the integrals computed for the two situations.

\subsubsection{$\gamma$ peak}

Since the $\gamma$ peak derives from the $d_{xy}$, $d_{yz}$. and $d_{xz}$ orbitals, we need to solve the generalized eigenvalue equation given in Eq.~(\ref{eq:g_eigen}), taking into account all three states on each adatom. However, from Eq.~(\ref{eq:Smat}), we see that the $d_{xy}$, $d_{yz}$, and $d_{xz}$ states decouple from each other, with
\begin{equation}
	S_{xz,xz}, C_{xz,xz}, D_{xz,xz} = S_2,C_2,D_2,
\end{equation}
and
\begin{equation}
	S_{xy,xy}, C_{xy,xy}, D_{xy,xy} =S_{yz,yz}, C_{yz,yz}, D_{yz,yz}= S_1,C_1,D_1.
\end{equation}
Hence, we can directly apply Eqs.~(\ref{eq:Epm_full}) and (\ref{eq:Epm_appr}) to compute the bound-state energy of the dimer. At large distance, we have
\begin{gather}
	E^{\gamma}_{xz,\pm} = E_{s}^{xy,yz,xz} + C^{\gamma}_2 \pm D^{\gamma}_{2}, \\
	E^{\gamma}_{xy,\pm} = E^{\gamma}_{yz,\pm} = E_{s}^{xy,yz,xz} + C^{\gamma}_1 \pm D^{\gamma}_{1}.
\end{gather}

\begin{figure}[h]
	\centering
	\includegraphics[width=0.7\textwidth]{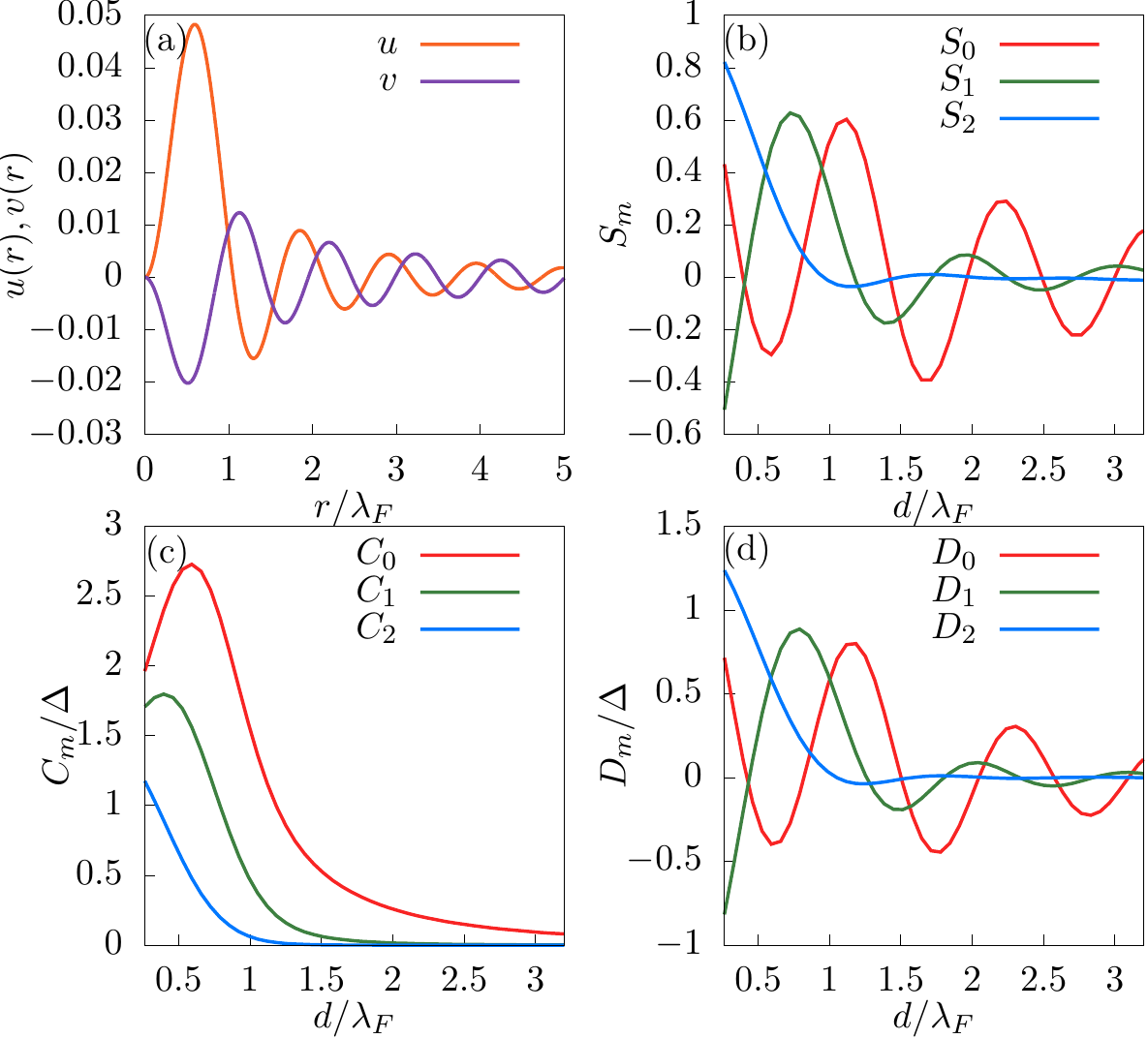}
	\caption{\label{fig:integrals}(a) The radial part of the Shiba state wave function. The electron and hole components
  are denoted as $u(r)$ and $v(r)$ respectively. (b--d) Overlap, Coulomb-like, and exchange-like integrals
  in terms of complex spherical harmonics with  magnetic quantum number $m$. }
\end{figure}

\begin{figure}[h]
	\centering
	\includegraphics[width=0.7\textwidth]{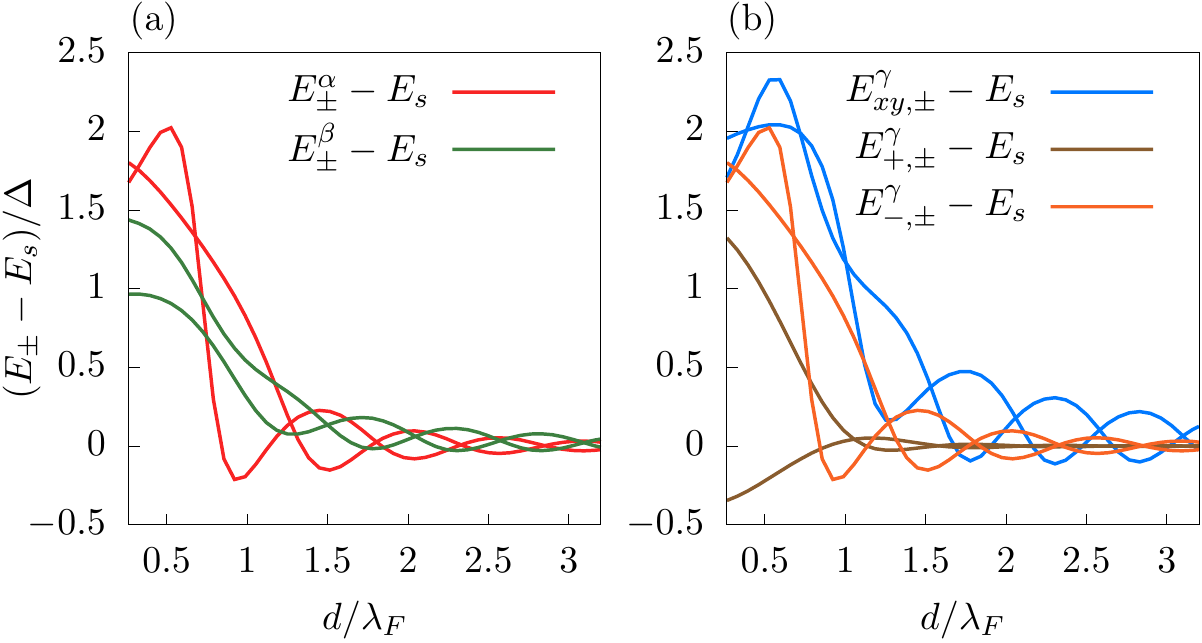}
	\caption{\label{fig:E110} The energy of Shiba states with two magnetic impurities oriented along \udir{110} direction, 
  measured from the Shiba state energy of an isolated impurity, originated from different $d$ orbitals.
  These states can be identified as $\alpha$, $\beta$ and $\gamma$ peaks according to the STM measurement.
  (a) $\alpha,\beta$ peaks. (b) $\gamma$ peak. }
\end{figure}

\begin{figure}[h]
	\centering
	\includegraphics[width=0.7\textwidth]{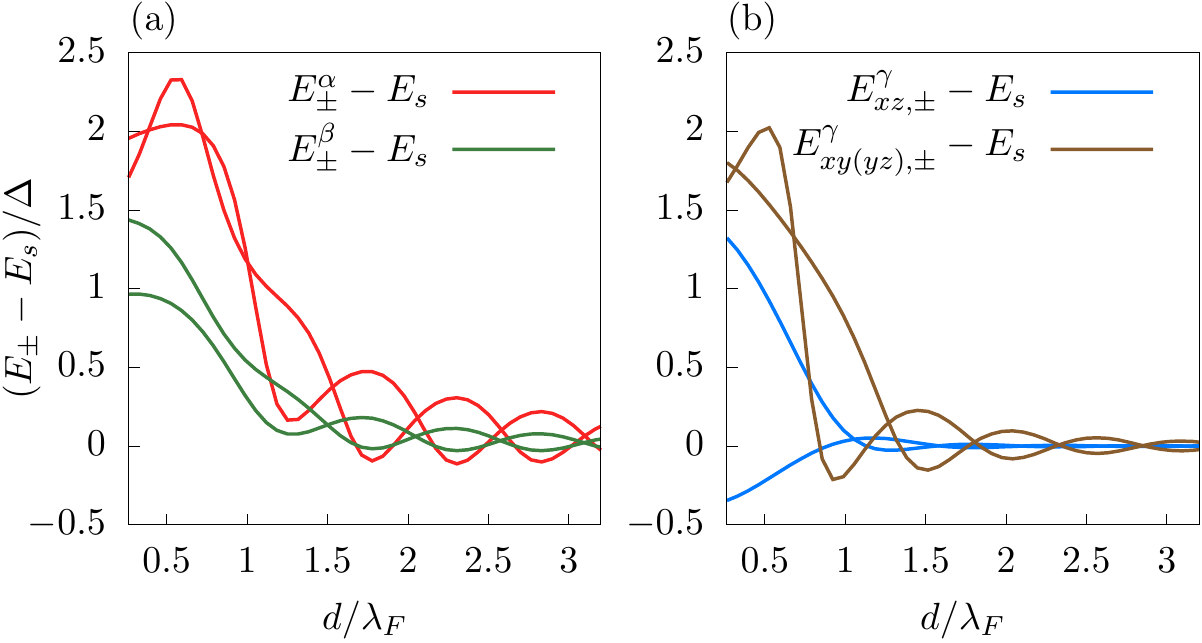}
	\caption{\label{fig:E100} The energy of Shiba states with two magnetic impurities oriented along \udir{100} direction, 
  measured from the Shiba state energy of an isolated impurity, originated from different $d$ orbitals.
  These states can be identified as $\alpha$, $\beta$ and $\gamma$ peaks according to the STM measurement.
  (a) $\alpha,\beta$ peaks. (b) $\gamma$ peak. }
\end{figure}

\section{Numerical Results}

For illustration, we present some numerical result. A full numerical implementation for realistic parameters is too demanding in view of the large ratio between coherence length and Fermi wavelength. Since the dimer dimension is very small compared to the coherence length, we keep realistic values for the Fermi wavelength, but reduce the coherence length significantly (while leaving it larger than the Fermi wavelength) by choosing an unrealistically large gap $\Delta$.  Moreover, we also truncate $k$ such that
\begin{equation}
	\frac{\alpha_{k,l}^2}{2R^2}\leq E_F + \epsilon.
	\label{eq:Debye}
\end{equation}
The cutoff $\epsilon$ should be chosen large compared to $\Delta$, but in practice, we choose it of order $\Delta$, so that the necessary basis set does not become too large. As a result, our numerical calculations generate reasonable d-orbital-like Shiba wave functions within the superconducting gap whose hybridization can then be studied within the variational approximation discussed above. The calculations provide qualitative insights into the hybridization but do not suffice for quantitative predictions.

Specifically, we take an unrealistically large superconducting gap of $\Delta=500\umeV$, but the Fermi energy $E_F = 9470 \umeV$ for $\mathrm{Pb}$, corresponding to a Fermi wavelength of $\lambda_F = 3.99\uAA$. We require the radius $R$ of the finite simulation space defined in Eq.~(\ref{eq:rhokl}) large enough, such that the level spacing (at fixed angular momentum) due to the finite size quantization is much smaller than the superconducting gap, namely $R\gg\sqrt{\mu}/\Delta$. We choose $R=761\uAA$ in order to  fulfill this requirement. Furthermore, we choose the cutoff $\epsilon$ in Eq.~(\ref{eq:Debye}) to be $250\umeV$, and an exchange potential 
\begin{equation}
	J(r) = \frac{V}{\sqrt{\pi a}}\exp\left(-\frac{r^2}{a^2}\right),
\end{equation}
where $a$ and $V$ characterize the range and the strength of the potential. Note that in the limit $a\to 0$, $J(r) \to V\delta(r)$. We choose $a=1.59\uAA$ and $V=122000 \umeV$ in order to produce Shiba states in the $l=2$ sector with energy $E_s=0.4274\Delta$. 

In Fig.~\ref{fig:integrals}(a), we show electron and hole components of the radial Shiba state wave function, denoted as $u(r)$ and $v(r)$. In Figs.~\ref{fig:integrals}(b--d), we show  $S_m$, $C_m$  and $D_m$
for $m=0,1,2$,  which are used  in computing the Shiba states energies for two impurities. The energies of Shiba states with two magnetic impurities oriented along \udir{110} and \udir{100} directions are shown in Figs.~\ref{fig:E110} and \ref{fig:E100}, respectively.

\end{document}